\documentclass[]{spie}  

 \usepackage{graphicx}
 \usepackage{cite}
 \usepackage{amsmath,amssymb,amsfonts}
 \usepackage{algorithmic}
 \usepackage{graphicx}
 \usepackage{textcomp}
 \usepackage{xcolor}
 
 \usepackage[utf8]{inputenc}
 \usepackage{colortbl,array} 
 \usepackage{multirow,bigdelim}
 \usepackage{amsmath}
 \usepackage{placeins}
 \usepackage{stfloats}
 \usepackage{amsfonts}
 \usepackage{booktabs}
 \usepackage{multirow}
 \usepackage[utf8]{inputenc}
 \usepackage{graphicx}
 \usepackage{tabularx}
 \newcolumntype{L}{>{\raggedright\arraybackslash}X}
 
 \DeclareUnicodeCharacter{FB01}{fi}

\usepackage{amsmath,amsfonts,amssymb}
\usepackage{graphicx}
\usepackage[colorlinks=true, allcolors=blue]{hyperref}

\title{Attention-based CT Scan Interpolation for Lesion Segmentation of Colorectal Liver Metastases}

\author[a,b]{Mohammad Hamghalam}
\author[c]{Richard K. G. Do}
\author[a,d]{Amber L. Simpson}
\affil[a]{School of Computing, Queen’s University, Kingston, ON, Canada}
\affil[b]{Department of Electrical Engineering, Qazvin Branch, Islamic Azad University, Qazvin, Iran}
\affil[c]{Department of Radiology, Memorial Sloan Kettering Cancer Center, New York, NY, USA}
\affil[d]{Department of Biomedical and Molecular Sciences, Queen's University, Kingston, ON, Canada.}

\authorinfo{Further author information: (Send correspondence to Amber L. Simpson)\\Amber L. Simpson: E-mail: amber.simpson@queensu.ca, \\  Mohammad Hamghalam: E-mail: m.hamghalam@gmail.com\\  Richard K. G. Do: E-mail: dok@mskcc.org \\
https://doi.org/10.1117/12.2656072
}

\begin{document} 
\maketitle
	\begin{abstract}
	Small liver lesions common to colorectal liver metastases (CRLMs) are challenging for convolutional neural network (CNN) segmentation models, especially when we have a wide range of slice thicknesses in the computed tomography (CT) scans. Slice thickness of CT images may vary by clinical indication. For example, thinner slices are used for presurgical planning when fine anatomic details of small vessels are required. While keeping the effective radiation dose in patients as low as possible, various slice thicknesses are employed in CRLMs due to their limitations. However, differences in slice thickness across CTs lead to significant performance degradation in CT segmentation models based on CNNs. This paper proposes a novel unsupervised attention-based interpolation model to generate intermediate slices from consecutive triplet slices in CT scans. We integrate segmentation loss during the interpolation model's training to leverage segmentation labels in existing slices to generate middle ones. Unlike common interpolation techniques in CT volumes, our model highlights the regions of interest (liver and lesions) inside the abdominal CT scans in the interpolated slice.
	Moreover, our model's outputs are consistent with the original input slices while increasing the segmentation performance in two cutting-edge 3D segmentation pipelines. We tested the proposed model on the CRLM dataset to upsample subjects with thick slices and create isotropic volume for our segmentation model. The produced isotropic dataset increases the Dice score in the segmentation of lesions and outperforms other interpolation approaches in terms of interpolation metrics.
	%
	%
	%
	
\end{abstract}
\section{Introduction}
\begin{figure}[!t]
	\centering
	\includegraphics[width=.92\columnwidth]{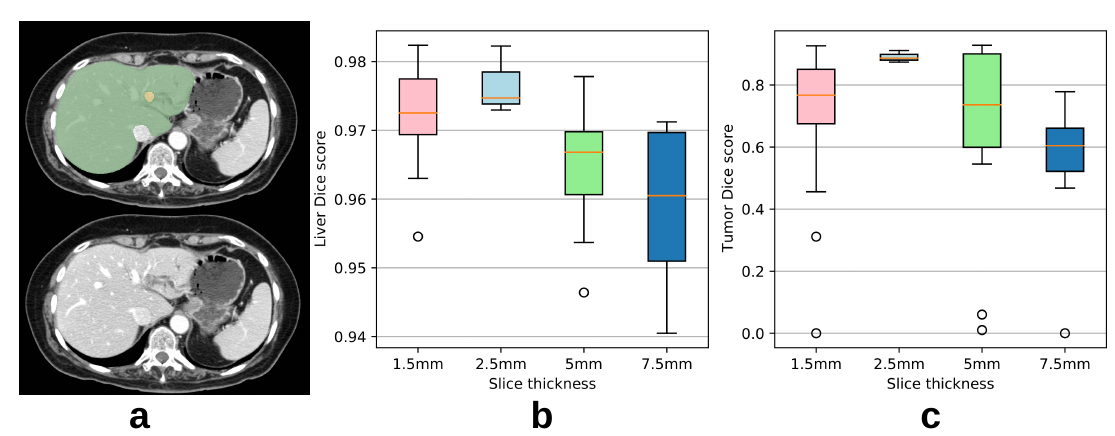} 
	\caption{Slice thicknesses of 1.5, 2.5, 5.0, and 7.5 $mm$ are available in our database, making small lesion segmentation challenging. (a) The 2D axial view of input abdominal CT scans and its segmentation map, including liver (green) and metastatic colorectal tumor (yellow) labels. The (b) and (c) are liver and tumor Dice scores versus slice thickness computed through the 3D UNet segmentation model via upsampling based on linear interpolation.}
	\label{fig1:slice_thickness}
\end{figure}

Computed tomography (CT) scan is the most commonly used imaging modality for patients with colorectal cancer, including those who develop liver metastases. CT is used to help identify the number and location of metastases, which has consequences for treatment options, including the possibility of surgical resection. In clinical practice, different CT protocols are used across and within institutions. Variations in protocols may include the choice of CT slice thickness, with consequences in detecting small liver metastases. The slice thickness is the distance between adjacent slices in the transverse (cross-sectional) plane of the patient. As the slice thickness increases, lesions smaller than the slice thickness may become more challenging to see. Conversely, very thin slices suffer from increased image noise, reducing the visibility of lesions from reductions in contrast to noise ratios. Fig. \ref{fig1:slice_thickness} (a) shows an example of colorectal liver metastases (CRLM) scans in the 2D axial view along with our regions of interest (ROIs), including liver and small lesion label.

Routine CT imaging protocols yield anisotropic voxels where the thickness between axial slices is typically greater than the cross-sectional resolution (i.e., $\Delta z > (\Delta x, \Delta y)$) \cite{whybra2019assessing}. It is advised to resample scans using 3D interpolation so that $\Delta z = \Delta x = \Delta y$ to maintain scale conservation in all three dimensions and eliminate a directional bias in 3D features. However, up-sampling to a smaller voxel size creates artificial information at a higher resolution, and extreme up-sampling makes the image look smoother and more homogeneous \cite{depeursinge2014three}.
While CT scan interpolation has been utilized for various reasons in the literature, we focus on acquiring isotropic voxel dimensions for the 3D convolutional neural network (CNN) segmentation models.

CNN and vision transformer models have achieved significant advances in the end-to-end 3D volumes segmentation\cite{mojtahedi2022towards,hatami2019machine,hamghalam2020brain,soleymanifard2019segmentation,hamghalam2020high,hamghalam2020convolutional,hamghalam2020transforming,hamghalam2020high,ghahremani2021local,hamghalam2021modality,soleymanifard2022multi,soleimany2017novel}, particularly UNet-based architectures \cite{ronneberger2015u,isensee2020nnu,hatamizadeh2022unetr}. However, typical slice thickness variation in CT imaging leads to significant segmentation degradation in the UNet-based models. Interpolation techniques have been applied to upsample low-resolution samples and create an isotropic dataset before feeding input images to the 3D UNet \cite{isensee2020nnu} and the UNETR \cite{hatamizadeh2022unetr} segmentation model. Though naive interpolation approaches address the various slice thickness problem in CT scans, linear and non-linear interpolation can result in severe artifacts and substantial loss of information in low-resolution CT volumes (Fig. \ref{fig1:slice_thickness} (b)). This issue will become more challenging when we need to segment small lesions in the CRLM subjects (Fig. \ref{fig1:slice_thickness} (c)). 

To increase resolution in 3D medical scans, Roger \textit{et al.} \cite{rogertowards} developed a model based on PixelCNN \cite{van2016conditional} that interpolates CT imaging slices better than standard interpolation techniques while maintaining quantitative imaging information. Chen \textit{et al.} \cite{chen2018efficient} presented a 3D model based on a generative adversarial network (GAN) \cite{goodfellow2014generative} to synthesize realistic output. They reached notable improvement compared to the nearest neighbour (NN) and bicubic interpolation in MRI scans. Taheri \textit{et al.} proposed a technique using both variational autoencoders and GAN to produce frames that are robust against heart rate variations in echocardiography.

In computer vision applications, Liu \textit{et al.} \cite{liu2017video} trained an encoder-decoder model based on triplets of consecutive video frames named deep voxel flow (DVF). Instead of fabricating pixel values of the target video frame \cite{mathieu2015deep}, the DVF directly learns the uniform motion for interpolating the middle frame. Bahat \textit{et al.} \cite{bahat2020explorable} designed an explorable interpolation model to produce several high-resolution images that dramatically vary in their textures and fine details from a low-resolution image.
However, the current interpolation models are restricted by the model’s disability to attend a specific region or tissue of input low-resolution image.

This paper presents a novel unsupervised interpolation model focusing on small liver lesions inside abdominal CT volumes in CRLM patients. The contribution of this work is three-fold. First, we integrate segmentation loss with the interpolation loss to provide attention to small liver lesions in the interpolated intermediate slices. Second, due to the lack of segmentation labels in the interpolated slices, we optimize our model to minimize the Dice score loss for center reconstructed slice, achieved by interpolating back from interpolated intermediate slices. Third, we validate our model on a clinical CRLM dataset, including a wide range of slice thickness. We resample our low-resolution data in the z-axis with the proposed interpolation model and feed to the 3D UNet model to segment our small liver lesions. The experiments confirm that incorporating the interpolated intermediate slices in the segmentation framework improves the average Dice score in detecting small metastasis lesions.

\section{Method}
\begin{figure}[!t]
	\centering
	\includegraphics[width=.92\columnwidth]{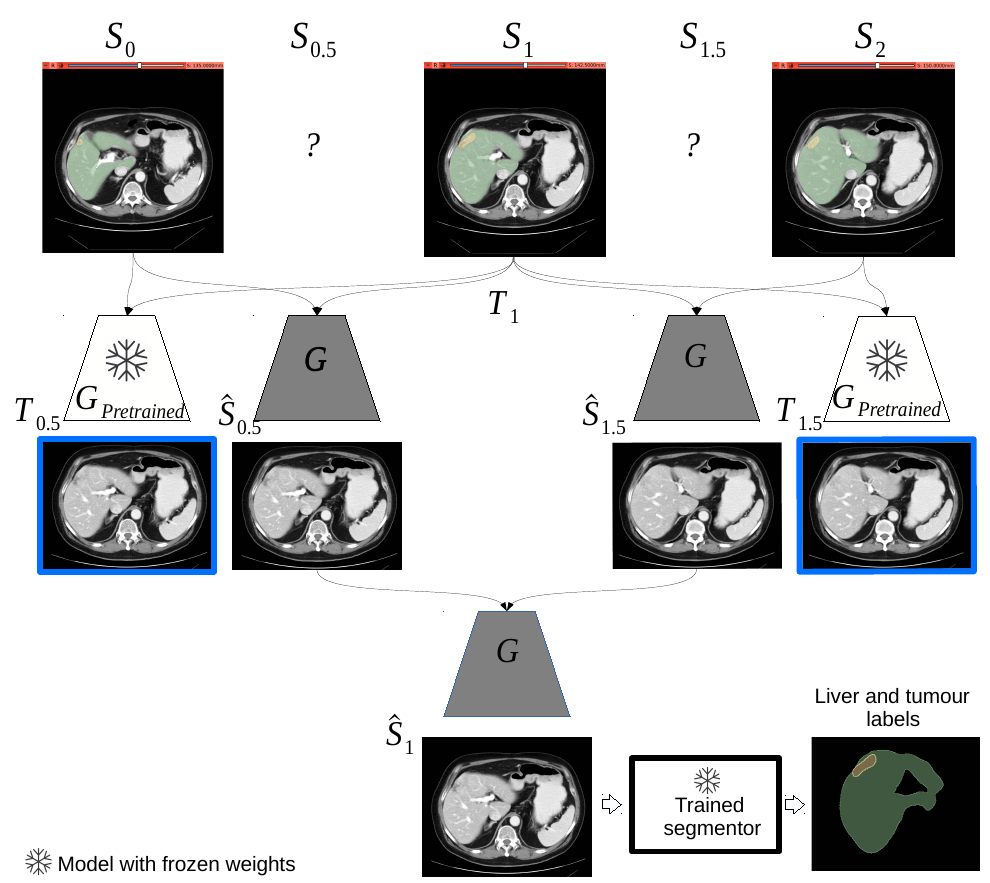} 
	\caption{Overview of the proposed interpolation method. The $\textbf{S}$ and $\hat{\textbf{S}}$ determine the original input slices and the predicted intermediate ones by our model. $T$ is GT generated through the pre-trained interpolation model. We incorporated segmentation loss in our training process to provide attention to liver and its lesions while generating intermediate slices.}
	\label{fig:overview}
\end{figure}

%
%
%
%
Assume we have three consecutive axial slices, $\textbf{S}_{0}$, $\textbf{S}_{1}$, and $\textbf{S}_{2}$, of abdominal CT scans as input with voxel spacing $5 mm$ in the scan direction (z-axis). Each slice is coupled with a segmentation map annotated by the radiologists. We propose to learn to interpolate intermediate slices, $\hat{\textbf{S}}_{n}$ and $\hat{\textbf{S}}_{1+n}$, where $n \in (0, 1)$, from available ones and their segmentation maps. As illustrated in Fig. \ref{fig:overview}, suppose that we need $2\textbf{x}$ upsampling in the z-axis, interpolating only one slice at $2.5 mm$. Thus, we have:
\begin{equation}\label{eq1}
	\begin{aligned}
		\hat{\textbf{S}}_{0.5}=G(\textbf{S}_{0}, \textbf{S}_{1}, n=0.5),\\
		\hat{\textbf{S}}_{1.5}=G(\textbf{S}_{1}, \textbf{S}_{2}, n=1.5)\\	
	\end{aligned}
\end{equation}
where $G$ is our slice interpolator with the encoder-decoder architecture. To provide interpolator attention to regions of interest (liver and its lesions), we define the segmentation loss, $\mathcal{L}_{Seg.}$, for the reconstructed center slice, $\hat{{\textbf{S}}}_{1}$, achieved by interpolating back from interpolated intermediate slices, $\hat{{\textbf{S}}}_{0.5}$ and $\hat{{\textbf{S}}}_{1.5}$, that is $\hat{\textbf{S}}_{1} = G(\hat{\textbf{S}}_{0.5}, \hat{\textbf{S}}_{1.5}, 1)$.  
Specifically, we feed $\hat{\textbf{S}}_{1}$ to a pre-trained segmentation UNet  model to detect liver and tumours pixels. This unsupervised model training is because of the absence of segmentation labels in interpolated slices.


The cycle consistency constraint is also defined in such a way that the resulting intermediate slice, $\hat{\textbf{S}}_{1}$, must match the original middle input slice $\textbf{S}_{1}$. So, we have:
%
%
%
%
\begin{equation}\label{eq2}
	\begin{aligned}	
		\mathcal{L}_{cycle} = {\left\lVert \hat{\textbf{S}}_{1}-{\textbf{S}}_{1} \right\rVert}_{1}
	\end{aligned}
\end{equation}	
%
%
%
%
%
For faster training coverage, we further constrain $G$ to resemble predictions of ${G}_{pre.}$ by pseudo supervised loss \cite{reda2019unsupervised}, $\mathcal{L}_{PS}$, as:

\begin{equation}\label{eq3}
	\begin{aligned}
		\mathcal{L}_{PS}=\left\lVert \hat{\textbf{S}}_{0.5}-{G}_{pre.}(\textbf{S}_{0}, \textbf{S}_{1}, n=0.5) \right\rVert_{1} + \left\lVert \hat{\textbf{S}}_{1.5}-{G}_{pre.}(\textbf{S}_{1}, \textbf{S}_{2}, n=1.5) \right\rVert_{1}
	\end{aligned}
\end{equation}
where ${G}_{pre.}$ has the same architecture as $G$ but is trained on a high-resolution disjoint CT dataset with voxel spacing $2.5 mm$ in the z-axis, which has ground-truth intermediate slices. Lastly, we regularize the contribution of each loss in our final objective function by $\lambda$, as:
\begin{equation}\label{eq4}
	\begin{aligned}	
		\mathcal{L} = {\lambda}_{cycle} \mathcal{L}_{cycle} + {\lambda}_{Seg.} \mathcal{L}_{Seg.} + {\lambda}_{PS} \mathcal{L}_{PS}
	\end{aligned}
\end{equation}
Fig. \ref{fig:overview} shows an overview of the proposed method for synthesizing interpolate slices.

\subsection{Slice interpolation using deep voxel flow}
To generate the intermediate slice from a pair of input slices, we designed $G$ inspired by DVF architecture that estimates voxel flow between slices. The voxel flow field is similar to an optical flow field and determines the relative movement between the objects and the scanner in consecutive slices. 
Specifically, the $G$ block has an encoder-decoder module followed by the trilinear interpolation block. The former consists of three convolution layers, three deconvolution layers and one bottleneck layer to produce voxel flows, $\textbf{F}$, while the latter synthesis intermediate slices based on the estimated voxel flows and existing slices. 
We trained $G_{pre.}$ on triplets of consecutive high-resolution CT slices with voxel spacing of $2.5 mm$ in the out-of-plane direction. To this end, we drop middle slices from the training set and employ the DVF loss function that measures similarity between generated pixels and the ground-truth (GT) dropped slices, $\textbf{S}_{0.5}$, as:
\begin{equation}\label{eq4}
	\begin{aligned}
		\mathcal{L}_{DVF}=\left\lVert \hat{\textbf{S}}_{0.5}-\textbf{S}_{0.5} \right\rVert_{1} + \left\lVert \nabla \textbf{F}_{motion} \right\rVert_{1} +  \left\lVert \nabla \textbf{F}_{mask} \right\rVert_{1}
	\end{aligned}
\end{equation}
where $\nabla \textbf{F}_{motion}$ is the variation term on in-plane direction components of voxel flow; similarly, $\nabla \textbf{F}_{mask}$ is the regularizer on in-plane direction of the voxel flow. The $G_{pre.}$ learns complex connections between existing slices by training on high-resolution CT scans and using dropped slices as GT.

\section{Experiments and Results}	
Our implementation is developed
employing Pytorch on an NVIDIA GeForce RTX 3090 GPU with
24G of RAM. Peak signal to noise (PSNR) and structural similarity metric (SSIM) have been used to evaluate slice interpolations quantitatively. High scores in both metrics indicate better quality. We also evaluate our segmentation pipeline by 4-fold cross-validation in terms of Dice score, recall, precision, and the average symmetric surface distance (ASD).	


\subsection{Datasets}
This research included 198 patients who underwent resection for CRLM with appropriate preoperative CT scans with a wide range of slice thicknesses of 1.5, 2.5, 5.0, 7.5 $mm$. Following standard imaging procedures, patients underwent a contrast-enhanced portal venous-phase CT scan \cite{simpson2017computed}. To train our $G {pre.}$ model, we also used the liver tumor segmentation (LiTS) dataset \cite{bilic2019liver}. In particular, we found 21 subjects (4394 slices) with a slice thickness of about $2.5 mm$ from LiTS and combined with 14 subjects (1296 slices) from the CRLM with $2.5 mm$ resolution. We divided these slices for training (0.8) and testing (0.2) our interpolation model.
For a fair comparison, as our interpolation model leverages the $G_{pre.}$ model pretrained on 21 subjects from the LiTS dataset, we integrated those samples with CRLM data in the training of compared methods.

%
%
\subsection{Attention-based CT scan interpolation results}
To evaluate our method, we interpolate the center slices based on three consecutive CT slices. We compare our method with the NN method and linear interpolation, which are prevalent upsampling pre-processing techniques in the segmentation of anisotropic CT scans \cite{isensee2020nnu,hatamizadeh2022unetr}. We also compare our model with recent CNN-based interpolation models, such as DVF, our baseline interpolation, and PixelMiner \cite{whybra2019assessing}, the convolutional CT scan interpolation strategy. Table \ref{tab:tab1} shows that our model outperforms both commonly used and CNN-based interpolation techniques in terms of PSNR and SSIM.
We also evaluate the effect of proposed loss functions in our region of interest of abdominal CT scans. As shown in Table \ref{tab:tab1}, we increase the PSNR and SSIM in the liver region by adding $\mathcal{L}_{Seg.}$ in our pipeline. This confirms that our design can attend to the specific ROIs.
\begin{table*}[!t]
	\centering
	\caption{Interpolation metrics for single intermediate slice interpolation on the abdominal CT scans (test set). The higher the better.} 
	\label{tab:tab1}
	\begin{tabular}{@{\extracolsep{\fill}}c|c|c|c|c|c|c|c}
		\toprule[1pt] 
		\textbf{Interpolation}  &\multirow{2}{*}{\textbf{NN}} 
		&\multirow{2}{*}{\textbf{Linear}} & \multirow{2}{*}{\textbf{DVF}}&\textbf{Pixel-} & \multicolumn{3}{c}{\textbf{Proposed}} \\
		\cmidrule[\heavyrulewidth]{1-1}
		\cmidrule[\heavyrulewidth]{6-8}
		\textbf{Metrics(ROI)} &  & &&\textbf{Miner}  & $\mathcal{L}_{cycle}$ & $\mathcal{L}_{cycle}+\mathcal{L}_{PS}$& $\mathcal{L}_{cycle}+\mathcal{L}_{PS}+\mathcal{L}_{Seg.}$\\
		\midrule
		\textbf{PSNR}(Whole)   & 20.21  & 23.11 & 25.65 &26.81 &  26.34 &  27.97 & 28.51 \\
		\textbf{SSIM}(Whole)   & 0.5070 & 0.5933  & 0.7633&7813 & 0.8152 & 0.8211 & 0.8271  \\
		\midrule
		\textbf{PSNR}(Liver)   & 19.85  & 22.67 & 24.23 &25.98 &  25.83 &  27.41 & \textbf{29.91} \\
		\textbf{SSIM}(Liver)   & 0.5023 & 0.5878  & 0.7612 &0.8177 & 0.8152 & 0.8211 & \textbf{0.8365}  \\
		\midrule
		%
		%
		
		%
		%
		
	\end{tabular}
\end{table*}
\subsection{Lesion segmentation in CRLMs}	
We upsample our CRLM dataset by interpolating one and two slices in subjects with 5 mm and 7.5 mm slice thickness, respectively. Table \ref{tab:tab2} and Table \ref{tab:tab3} show the segmentation results by two state-of-the-art medical image segmentation pipelines, the 3D UNet and UNETR model, with both conventional and convolutional interpolation techniques to validate the proposed approach. Our attention-based interpolation method can improve segmentation performance, especially in small ROIs (lesions) compared to the liver.
For implementation, we employed the nnUNet (the 3D full resolution structure)\footnote{https://github.com/MIC-DKFZ/nnUNet} \cite{isensee2020nnu} and the UNETR\footnote{https://github.com/Project-MONAI/research-contributions/tree/master/UNETR} \cite{hatamizadeh2022unetr}, two cutting-edge segmentation pipelines, in these experiments. The latter is a transformer-based model with state-of-the-art CT scan segmentation performance, while the former has a 3D UNet architecture and achieves prominence on various medical image benchmarks \cite{antonelli2021medical}.
\begin{table*}[!h]
	\centering
	\caption{Liver segmentation results on CRLM dataset through 3D UNet and UNETR with different interpolation techniques.} 
	\label{tab:tab2}
	\begin{tabular}{@{\extracolsep{\fill}}c|c|c|c|c|c}
		\toprule[1pt] 
		\textbf{Interpolation} & \textbf{Segmentation Baseline} 
		&\textbf{DSC} & \textbf{Recall} & \textbf{Precision} & \textbf{ASD}($\downarrow$) \\
		
		\midrule
		\multirow{2}{*}{\textbf{NN}}  & 3D UNet & 0.9536 & 0.9867 & 0.9227 & 1.136 \\
		& UNETR & 0.8937 & 0.9178 & 0.9696 & 1.237 \\
		\midrule
		\multirow{2}{*}{\textbf{Linear}}  & 3D UNet & 0.9613 & 0.9909 & 0.9334 & 1.051 \\
		& UNETR & 0.9021 & 0.9154 & 0.8644 & 1.154 \\
		\midrule
		\multirow{2}{*}{\textbf{DVF}}& 3D UNet & 0.9659 & 0.9847 & 0.9478 & 1.089 \\
		& UNETR & 0.9081 & 0.9201 & 0.9152 & 1.002 \\
		\midrule
		\multirow{2}{*}{\textbf{PixelMiner}}& 3D UNet & 0.9677 & 0.9881 & 0.9513 & 1.065 \\
		& UNETR & 0.9105 & 0.9259 & 0.9188 & 1.001 \\
		\midrule		  
		\multirow{2}{*}{\textbf{Proposed (${\lambda}_{Seg.}=0$)}}  & 3D UNet & 0.9675 & 0.9685 & \textbf{0.9675 }& 1.01 \\
		& UNETR & 0.9211 & 0.9350 & 0.9238 & 0.955 \\
		\midrule
		\multirow{2}{*}{\textbf{Proposed (${\lambda}_{Seg.}\neq0$)}}  & 3D UNet & \textbf{0.9754} & \textbf{0.9941} & 0.9574 & \textbf{0.990} \\
		& UNETR & \textbf{0.9366} & \textbf{0.9591} &\textbf{ 0.9449} & \textbf{0.931} \\
		\midrule
	\end{tabular}
\end{table*}
\begin{table*}[!h]
	\centering
	\caption{Tumor segmentation results on CRLM dataset through 3D UNet and UNETR with different interpolation techniques.} 
	\label{tab:tab3}
	\begin{tabular}{@{\extracolsep{\fill}}c|c|c|c|c|c}
		\toprule[1pt] 
		\textbf{Interpolation} & \textbf{Segmentation Baseline} 
		&\textbf{DSC} & \textbf{Recall} & \textbf{Precision} & \textbf{ASD}($\downarrow$) \\
		
		\midrule
		\multirow{2}{*}{\textbf{NN}} & 3D UNet & 0.6658 & 0.6445 & 0.8275 & 1.170 \\		
		& UNETR  & 0.5094 & 0.4921 & 0.7412 & 1.341 \\
		
		\midrule
		\multirow{2}{*}{\textbf{Linear}}  &  3D UNet & 0.6818 & 0.5195 & 0.9914 & 1.081 \\
		& UNETR & 0.5366 & 0.4442 & 0.9012 & 1.250 \\
		
		\midrule
		\multirow{2}{*}{\textbf{DVF}}& 3D UNet & 0.7088 & 0.5504 & \textbf{0.9952} & 0.838 \\
		& UNETR & 0.5588 & 0.5033 & 0.9139 & 1.118 \\
		
		\midrule
		\multirow{2}{*}{\textbf{PixelMiner}}& 3D UNet& 0.7427 & 0.5611 & 0.9950 & 0.814 \\
		& UNETR & 0.5916 & 0.5710 & 0.9388 & 1.118 \\
		
		\midrule
		\multirow{2}{*}{\textbf{Proposed (${\lambda}_{Seg.}=0$)}}  & 3D UNet & 0.7383 & 0.5361 & 0.9944 & 0.825 \\
		& UNETR & 0.5861 & 0.5699 & 0.9340 & 1.010 \\
		
		\midrule
		\multirow{2}{*}{\textbf{Proposed (${\lambda}_{Seg.}\neq0$)}} & 3D UNet & \textbf{0.7782} & \textbf{0.6564} & 0.9555 & \textbf{0.781} \\
		&UNETR & \textbf{0.6154} & \textbf{0.5811} & \textbf{0.9474} & \textbf{0.980} \\
		\midrule
	\end{tabular}
\end{table*}
%

\begin{figure}[!h]
	\centering
	\includegraphics[width=.99\columnwidth]{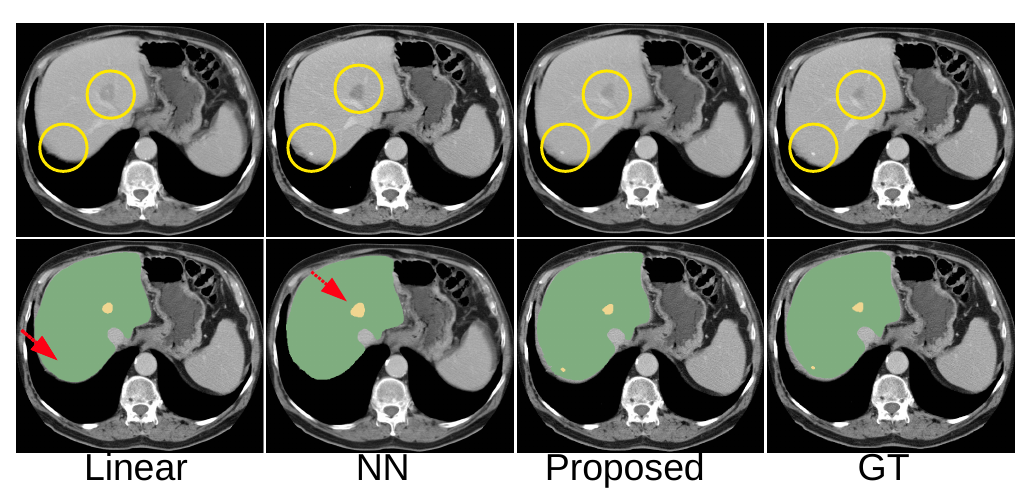} 
	\caption{Visual comparison of interpolation (first row) and segmentation (second row) on the CRLM dataset. The yellow label is the lesion, and the green one shows the liver in the axial view of abdominal scans. Zoom-in to view small lesions inside the circles.}
	\label{fig3:results}
\end{figure}

%
%
As shown in Fig. \ref{fig3:results}, in the linear interpolation model, the small lesions are lost in the interpolated slice. This fact decreases the segmentation accuracy. The NN model changes the tumor size in the generated intermediate image. The proposed method can detect the small and large lesions precisely in the interpolation. This is because of integrating the segmentation loss during training and penalize our model when we missed the small lesions. 	


\section{Discussion and conclusion}
The small lesions expected in CRLM are challenging for the CNN-based segmentation models, particularly when we have a wide range of slice thicknesses in our dataset. Various slice thickness is because of different CT protocols and patient conditions in abdominal scans. We introduced an attention-based interpolation model to upsample low-resolution subjects with attention to the liver and lesions. We combined the segmentation loss with the interpolation loss to draw attention to small liver lesions in the interpolated intermediate slices, which would otherwise be overlooked. Our unsupervised interpolation method precisely predicted tiny tumors and increased quality metrics in the liver region. We also significantly improved the DSC in the small lesion regions using our attention-based interpolation techniques compared to conventional interpolation approaches.

\acknowledgments

This work was funded in part by National Institutes of Health R01CA233888.

%
%
%
%
\bibliographystyle{spiebib}
\bibliography{mybib_miccai2020_abs}

\begin{thebibliography}{10}

\bibitem{whybra2019assessing}
Whybra, P., Parkinson, C., Foley, K., Staffurth, J., and Spezi, E., ``Assessing
  radiomic feature robustness to interpolation in 18f-fdg pet imaging,'' {\em
  Scientific reports}~{\bf 9}(1),  1--10 (2019).

\bibitem{depeursinge2014three}
Depeursinge, A., Foncubierta-Rodriguez, A., Van De~Ville, D., and M{\"u}ller,
  H., ``Three-dimensional solid texture analysis in biomedical imaging: review
  and opportunities,'' {\em Medical image analysis}~{\bf 18}(1),  176--196
  (2014).

\bibitem{mojtahedi2022towards}
Mojtahedi, R., Hamghalam, M., Do, R.~K., and Simpson, A.~L., ``Towards optimal
  patch size in vision transformers for tumor segmentation,'' in [{\em
  Multiscale Multimodal Medical Imaging: Third International Workshop, MMMI
  2022, Held in Conjunction with MICCAI 2022, Singapore, September 22, 2022,
  Proceedings}{\nolinebreak\hspace{0.1em}]},   110--120, Springer (2022).

\bibitem{hatami2019machine}
Hatami, T., Hamghalam, M., Reyhani-Galangashi, O., and Mirzakuchaki, S., ``A
  machine learning approach to brain tumors segmentation using adaptive random
  forest algorithm,'' in [{\em 2019 5th Conference on Knowledge Based
  Engineering and Innovation (KBEI)}{\nolinebreak\hspace{0.1em}]},   076--082,
  IEEE (2019).

\bibitem{hamghalam2020brain}
Hamghalam, M., Lei, B., and Wang, T., ``Brain tumor synthetic segmentation in
  3d multimodal mri scans,'' in [{\em Brainlesion: Glioma, Multiple Sclerosis,
  Stroke and Traumatic Brain Injuries: 5th International Workshop, BrainLes
  2019, Held in Conjunction with MICCAI 2019, Shenzhen, China, October 17,
  2019, Revised Selected Papers, Part I 5}{\nolinebreak\hspace{0.1em}]},
  153--162, Springer International Publishing (2020).

\bibitem{soleymanifard2019segmentation}
Soleymanifard, M. and Hamghalam, M., ``Segmentation of whole tumor using
  localized active contour and trained neural network in boundaries,'' in [{\em
  2019 5th Conference on Knowledge Based Engineering and Innovation
  (KBEI)}{\nolinebreak\hspace{0.1em}]},   739--744, IEEE (2019).

\bibitem{hamghalam2020high}
Hamghalam, M., Lei, B., and Wang, T., ``High tissue contrast mri synthesis
  using multi-stage attention-gan for glioma segmentation,'' in [{\em The
  Thirty-Fourth AAAI Conference on Artificial Intelligence
  (AAAI-20)}{\nolinebreak\hspace{0.1em}]},   {\bf 34}(04),  4067--4074 (2020).

\bibitem{hamghalam2020convolutional}
Hamghalam, M., Lei, B., and Wang, T., ``Convolutional 3d to 2d patch conversion
  for pixel-wise glioma segmentation in mri scans,'' in [{\em Brainlesion:
  Glioma, Multiple Sclerosis, Stroke and Traumatic Brain Injuries: 5th
  International Workshop, BrainLes 2019, Held in Conjunction with MICCAI 2019,
  Shenzhen, China, October 17, 2019, Revised Selected Papers, Part I
  5}{\nolinebreak\hspace{0.1em}]},   3--12, Springer International Publishing
  (2020).

\bibitem{hamghalam2020transforming}
Hamghalam, M., Wang, T., Qin, J., and Lei, B., ``Transforming intensity
  distribution of brain lesions via conditional gans for segmentation,'' in
  [{\em 2020 IEEE 17th international symposium on biomedical imaging
  (ISBI)}{\nolinebreak\hspace{0.1em}]},   1--4, IEEE (2020).

\bibitem{ghahremani2021local}
Ghahremani, M., Ghadiri, H., and Hamghalam, M., ``Local features integration
  for content-based image retrieval based on color, texture, and shape,'' {\em
  Multimedia Tools and Applications}~{\bf 80}(18),  28245--28263 (2021).

\bibitem{hamghalam2021modality}
Hamghalam, M., Frangi, A.~F., Lei, B., and Simpson, A.~L., ``Modality
  completion via gaussian process prior variational autoencoders for
  multi-modal glioma segmentation,'' in [{\em Medical Image Computing and
  Computer Assisted Intervention--MICCAI 2021: 24th International Conference,
  Strasbourg, France, September 27--October 1, 2021, Proceedings, Part VII
  24}{\nolinebreak\hspace{0.1em}]},   442--452, Springer International
  Publishing (2021).

\bibitem{soleymanifard2022multi}
Soleymanifard, M. and Hamghalam, M., ``Multi-stage glioma segmentation for
  tumour grade classification based on multiscale fuzzy c-means,'' {\em
  Multimedia Tools and Applications}~{\bf 81}(6),  8451--8470 (2022).

\bibitem{soleimany2017novel}
Soleimany, S. and Hamghalam, M., ``A novel random-valued impulse noise detector
  based on mlp neural network classifier,'' in [{\em 2017 Artificial
  Intelligence and Robotics (IRANOPEN)}{\nolinebreak\hspace{0.1em}]},
  165--169, IEEE (2017).

\bibitem{ronneberger2015u}
Ronneberger, O. et~al., ``U-net: Convolutional networks for biomedical image
  segmentation,'' in [{\em MICCAI}{\nolinebreak\hspace{0.1em}]},   234--241
  (2015).

\bibitem{isensee2020nnu}
Isensee, F., Jaeger, P.~F., Kohl, S.~A., Petersen, J., and Maier-Hein, K.~H.,
  ``nnu-net: a self-configuring method for deep learning-based biomedical image
  segmentation,'' {\em Nature Methods} ,  1--9 (2020).

\bibitem{hatamizadeh2022unetr}
Hatamizadeh, A., Tang, Y., Nath, V., Yang, D., Myronenko, A., Landman, B.,
  Roth, H.~R., and Xu, D., ``Unetr: Transformers for 3d medical image
  segmentation,'' in [{\em Proceedings of the IEEE/CVF Winter Conference on
  Applications of Computer Vision}{\nolinebreak\hspace{0.1em}]},   574--584
  (2022).

\bibitem{rogertowards}
Roger, W., Lambin, P., Keek, S., and thers, ``Towards texture accurate slice
  interpolation of medical images using pixelminer,'' {\em
  10.21203/rs.3.rs-586453/v1}  (2021).

\bibitem{van2016conditional}
Van~den Oord, A., Kalchbrenner, N., Espeholt, L., Vinyals, O., Graves, A.,
  et~al., ``Conditional image generation with pixelcnn decoders,'' {\em
  Advances in neural information processing systems}~{\bf 29} (2016).

\bibitem{chen2018efficient}
Chen, Y., Shi, F., Christodoulou, A.~G., Xie, Y., Zhou, Z., and Li, D.,
  ``Efficient and accurate mri super-resolution using a generative adversarial
  network and 3d multi-level densely connected network,'' in [{\em
  International Conference on Medical Image Computing and Computer-Assisted
  Intervention}{\nolinebreak\hspace{0.1em}]},   91--99, Springer (2018).

\bibitem{goodfellow2014generative}
Goodfellow, I.~J., Pouget-Abadie, J., Mirza, M., Xu, B., Warde-Farley, D.,
  Ozair, S., Courville, A., and Bengio, Y., ``Generative adversarial
  networks,'' {\em arXiv preprint arXiv:1406.2661}  (2014).

\bibitem{liu2017video}
Liu, Z., Yeh, R.~A., Tang, X., Liu, Y., and Agarwala, A., ``Video frame
  synthesis using deep voxel flow,'' in [{\em Proceedings of the IEEE
  International Conference on Computer Vision}{\nolinebreak\hspace{0.1em}]},
  4463--4471 (2017).

\bibitem{mathieu2015deep}
Mathieu, M., Couprie, C., and LeCun, Y., ``Deep multi-scale video prediction
  beyond mean square error,'' {\em ICLR}  (2016).

\bibitem{bahat2020explorable}
Bahat, Y. and Michaeli, T., ``Explorable super resolution,'' in [{\em
  Proceedings of the IEEE/CVF Conference on Computer Vision and Pattern
  Recognition}{\nolinebreak\hspace{0.1em}]},   2716--2725 (2020).

\bibitem{reda2019unsupervised}
Reda, F.~A., Sun, D., Dundar, A., Shoeybi, M., Liu, G., Shih, K.~J., Tao, A.,
  Kautz, J., and Catanzaro, B., ``Unsupervised video interpolation using cycle
  consistency,'' in [{\em Proceedings of the IEEE/CVF International Conference
  on Computer Vision}{\nolinebreak\hspace{0.1em}]},   892--900 (2019).

\bibitem{simpson2017computed}
Simpson, A.~L., Doussot, A., Creasy, J.~M., Adams, L.~B., Allen, P.~J.,
  DeMatteo, R.~P., G{\"o}nen, M., Kemeny, N.~E., Kingham, T.~P., Shia, J.,
  et~al., ``Computed tomography image texture: a noninvasive prognostic marker
  of hepatic recurrence after hepatectomy for metastatic colorectal cancer,''
  {\em Annals of surgical oncology}~{\bf 24}(9),  2482--2490 (2017).

\bibitem{bilic2019liver}
Bilic, P., Christ, P.~F., Vorontsov, E., Chlebus, G., Chen, H., Dou, Q., Fu,
  C.-W., Han, X., Heng, P.-A., Hesser, J., et~al., ``The liver tumor
  segmentation benchmark (lits),'' {\em arXiv preprint arXiv:1901.04056}
  (2019).

\bibitem{antonelli2021medical}
Antonelli, M., Reinke, A., Bakas, S., et~al., ``The medical segmentation
  decathlon,'' {\em arXiv preprint arXiv:2106.05735}  (2021).

\end{thebibliography}

\end{document}